\documentclass[pre,twocolumn,superscriptaddress,amsmath,amssymb,floatfix]{revtex4-2}

\usepackage{graphicx}
\usepackage{bm}
\usepackage[colorlinks=true,allcolors=blue]{hyperref}

\newcommand{\x}{\mathbf{x}}

\newcommand{\dd}{\mathrm{d}}

\begin{document}

\title{Mate-finding Allee and crowding terms in replicator dynamics}

\author{Marcelo N. Kuperman}
\affiliation{Centro At\'omico Bariloche and Instituto Balseiro, Comisi\'on Nacional de Energ\'{\i}a At\'omica, 8400 San Carlos de Bariloche, R\'{\i}o Negro, Argentina}
\affiliation{Consejo Nacional de Investigaciones Cient\'{\i}ficas y T\'ecnicas (CONICET), Argentina}

\begin{abstract}
We study a simple model of replicator dynamics in which the fitness of each strategy is its game payoff plus a per-capita demographic rate that combines a mate-finding Allee effect with logistic crowding. We derive the frequency equations exactly from strategy-specific demography at constant total density, where the mortality and the Allee threshold disappear from the frequency equations. The demographic term shifts all the eigenvalues of the symmetric state by the same amount, and it can destabilize that state only above a critical strength of the Allee effect, which decreases with the number of strategies. For two strategies the term acts as a rarity advantage, which comes from the type-specific crowding and not from mate limitation, and which produces bistable coexistence of a cooperator minority in the prisoner's dilemma and tristability in coordination games. In rock-paper-scissors we obtain a Hopf curve and a Bautin point in closed form, and we find stable and unstable isolated limit cycles that the linear replicator equation cannot produce. When the total density is free, or when the types share the crowding at a density below the carrying capacity, the demographic term penalizes the invasion of the pure states and widens the range of payoffs for which the heteroclinic cycle attracts the orbits.
\end{abstract}

\maketitle

\section{Introduction}

Most theoretical studies describe the evolution of strategies in a population through the replicator equation~\cite{TaylorJonker1978,HofbauerSigmund1998,HofbauerSigmund2003,Nowak2006book}, where the frequency of each strategy grows or decays according to the difference between its fitness and the average fitness. In its classical form the fitness is linear in the frequencies, $f_i=(\mathsf{A}\x)_i$, and the literature offers a complete classification of its phase portraits for two and three strategies~\cite{Zeeman1980,HofbauerSigmund1998}. This linear setting has some well known rigidities. In two-strategy games the interior equilibrium, when it exists, is unique. In rock-paper-scissors (RPS) the interior equilibrium undergoes a degenerate Hopf bifurcation, and isolated periodic orbits cannot exist~\cite{Zeeman1980}. Any departure from linearity can therefore change the dynamics in a qualitative way, as several works have shown for multiplayer games~\cite{GokhaleTraulsen2010,Pacheco2009}, higher-order interactions~\cite{GriffinWu2023}, mutations~\cite{Mobilia2010,ToupoStrogatz2015}, environmental feedback~\cite{Weitz2016}, and explicit population dynamics~\cite{Hauert2006,GokhaleHauert2016,Krivan2018}.

Among the demographic nonlinearities, the Allee effect has a clear biological relevance. It refers to a positive relation between per-capita growth and density at low densities, both in single populations and in interacting species~\cite{Allee1931,Stephens1999,Courchamp2008,Kuperman2021}, and one of its best documented mechanisms is mate limitation: the probability that an individual finds a mate saturates with the density of conspecifics as $N/(N+\theta)$~\cite{Dennis1989,Boukal2002}. If we combine this factor with logistic crowding, the per-capita birth rate becomes $\propto N(1-N/K)/(N+\theta)$, a hump-shaped function of the density.

The closest antecedent to the present work is the study of Dhiman and Poria~\cite{Dhiman2018}, who introduced an Allee-type growth function $g(x)=rx(x/x_A-1)(1-x/x_K)$ into the generalized replicator equation of Wesson and Rand~\cite{WessonRand2013}, $\dot x_i=g(x_i)(f_i-\phi)$, with $\phi=\sum_kg(x_k)f_k/\sum_kg(x_k)$. For zero-sum RPS they found 31 equilibrium points, 13 of which can be stable. In that formulation $g$ multiplies the selection term instead of entering the fitness. For $0<x_i<x_A$ it reverses the direction of selection, the additional stable states appear where a frequency reaches a zero of $g$ and stays there regardless of payoff differences, and the interior equilibrium keeps purely imaginary eigenvalues for all parameters. To our knowledge, it remains open which game classes an Allee term derived from demography affects, and through which bifurcations.

For a single population with density $u$ in units of $K$, the mate-finding mechanism gives
\begin{equation}
\dot u=u\big[\varphi(u)-m\big],\qquad \varphi(u)=\frac{u(1-u)}{u+a},
\label{eq:allee1d}
\end{equation}
where $a=\theta/K>0$ is the mate-finding half-saturation constant in units of the carrying capacity, and $m>0$ is a density-independent mortality. A small $a$ corresponds to a weak Allee effect (individuals find mates easily) and a large $a$ to a strong one. The nontrivial fixed points of Eq.~\eqref{eq:allee1d} solve $u^2-(1-m)u+am=0$:
\begin{equation}
u_{A,C}=\tfrac12\Big[(1-m)\mp\sqrt{(1-m)^2-4am}\Big].
\label{eq:uAC}
\end{equation}
These points exist provided $(1-m)^2>4am$. The extinct state $u=0$ is stable, $u_A$ is the unstable Allee threshold, and $u_C$ is the stable population size. This is the structure of a strong Allee effect~\cite{Courchamp2008,Kumar2009}, and the mortality $m$ creates the threshold, since $u_A=0$ for $m=0$.

In this work we present an analysis of the replicator equation in which the fitness of each strategy is this per-capita rate plus the game payoff,
\begin{equation}
f_i(\x)=\varphi(x_i)-m+(\mathsf{A}\x)_i .
\label{eq:fitness}
\end{equation}
It is not our intention to describe a particular biological system. On the contrary, our more modest goal is to show which qualitative changes a single demographic term produces in the simplest games.

We first obtain Eq.~\eqref{eq:fitness} exactly from strategy-specific demography at constant total density, where the mortality and the Allee threshold disappear from the frequency equations (Sec.~\ref{sec:model}). We then show that the self-term leaves the invasion exponents at the pure states unchanged, penalizes the invasion of mixed populations, and shifts the spectrum of the symmetric state by an amount whose sign depends on whether the strategies occupy the Allee or the crowding branch of $\varphi$ (Sec.~\ref{sec:general}). For two strategies the self-term acts as a rarity advantage, and a concavity argument gives the complete count of interior equilibria (Sec.~\ref{sec:two}). In RPS we obtain the Hopf curve and its first Lyapunov coefficient in closed form, together with a Bautin point, a fold of cycles, and a criterion for the stability of the heteroclinic cycle, and these elements produce stable and unstable isolated limit cycles (Sec.~\ref{sec:rps}). Finally, we consider a variable total density (Sec.~\ref{sec:freeN}) and a crowding shared among the types (Sec.~\ref{sec:crowd}). In both cases the pure states become harder to invade, and in the second one the rarity advantage survives only above a critical weight of the type-specific crowding.

\section{Model}
\label{sec:model}

\subsection{Demographic derivation}

We consider $n$ types (strategies, morphs, or strains) with densities $N_i$. Each type reproduces within its own mating pool and experiences its own crowding, which we can interpret as the use of a type-specific niche or resource. Each type also receives a payoff from random encounters with the whole population. In units in which the maximal birth rate equals one, the per-capita growth rate reads
\begin{equation}
\frac{\dot N_i}{N_i}=\varphi\!\left(\frac{N_i}{K}\right)-m+(\mathsf{A}\x)_i-D(t),\qquad x_i=\frac{N_i}{N},
\label{eq:demo}
\end{equation}
with $N=\sum_k N_k$, $m$ a density-independent mortality, and $\mathsf{A}=(A_{ij})$ the payoff matrix. The last term $D(t)$ represents an external regulation that keeps the total density constant, as the dilution in a turbidostat~\cite{SmithWaltman1995} or the constant-size constraint of Moran-type models. The condition $\dot N=0$ gives $D=\sum_k x_k[\varphi(N_k/K)-m+(\mathsf{A}\x)_k]$. If the fixed density equals the carrying capacity, $N\equiv K$, then $N_i/K=x_i$ and the frequencies obey
\begin{equation}
\dot x_i=x_i\big[f_i(\x)-\bar f(\x)\big],\qquad \bar f=\sum_k x_k f_k,
\label{eq:rep}
\end{equation}
with $f_i$ given by Eq.~\eqref{eq:fitness}. Since $-m$ enters every $f_i$ in the same way, it cancels out in $f_i-\bar f$. Type-specific mortalities $m_i$ do not cancel, but we can absorb them into $\mathsf{A}$ by adding $-m_i$ to the $i$th row, since $\sum_j x_j=1$. The Allee \emph{threshold} $u_A$ of Eq.~\eqref{eq:uAC}, which exists only because $m>0$, thus plays no role at constant density. It reappears when $N$ is free (Sec.~\ref{sec:freeN}). From here on we use the equivalent fitness $f_i=\varphi(x_i)+(\mathsf A\x)_i$.

The function $\varphi$ has a hump shape [Fig.~\ref{fig:phi}(a)]. It increases for $u<u_m(a)$ (Allee branch) and decreases for $u>u_m(a)$ (crowding branch), with
\begin{equation}
u_m(a)=\sqrt{a^2+a}-a\in(0,\tfrac12),\qquad \varphi(0)=\varphi(1)=0.
\label{eq:um}
\end{equation}
Since $\varphi$ is smooth for $a>0$, Eq.~\eqref{eq:rep} leaves the simplex $\Delta^{n-1}$ and all its faces invariant~\cite{HofbauerSigmund1998}.

\section{General properties}
\label{sec:general}

At the vertex $\mathbf{e}_i$ the transversal eigenvalues of Eq.~\eqref{eq:rep} are
\begin{equation}
\lambda_{j|i}=f_j(\mathbf{e}_i)-f_i(\mathbf{e}_i)=A_{ji}-A_{ii},
\label{eq:vertex}
\end{equation}
exactly as in the linear replicator equation, since $\varphi(0)=\varphi(1)=0$. The Allee term therefore does not affect the Nash or ESS character of any pure strategy, nor the linear stability of heteroclinic cycles between vertices, whenever they exist.

The situation is different for a mixed resident state $\x$ on a face with $x_j=0$. The invasion exponent of strategy $j$ is
\begin{equation}
\begin{split}
\lambda_j(\x)&=(\mathsf{A}\x)_j-\x^{\!\top}\mathsf{A}\x-\Psi(\x),\\
\Psi(\x)&=\sum_k x_k\varphi(x_k)\ge0,
\end{split}
\label{eq:penalty}
\end{equation}
with $\Psi=0$ only at the vertices. This penalty is the signature of the type-specific self-term at the level of frequencies. The residents benefit from the abundance of conspecifics, while an absent strategy does not. A mixed resident population is thus harder to invade than in the linear game, whereas a pure resident population is not.

At the barycenter $\bar\x=(1/n)\mathbf{1}$ we can compute the effect of the self-term exactly when $\bar\x$ is an equilibrium of the linear game with $\mathsf{A}\bar\x\propto\mathbf{1}$ and $\mathsf{A}^{\!\top}\bar\x\propto\mathbf{1}$, as for circulant matrices. Then $\bar\x$ is also an equilibrium of Eq.~\eqref{eq:rep}. If we linearize $x_i[\varphi(x_i)-\sum_kx_k\varphi(x_k)]$ at $\bar\x$ along a vector $\mathbf v$ with $\sum_iv_i=0$, we obtain $\frac1n[\varphi'(\frac1n)v_i-\sum_k(\varphi(\frac1n)+\frac1n\varphi'(\frac1n))v_k]=\frac1n\varphi'(\frac1n)v_i$, while the game terms coincide with those of the linear replicator, since $\mathsf{A}\bar\x$ and $\mathsf{A}^{\!\top}\bar\x$ are parallel to $\mathbf 1$. On the tangent space the Jacobian is then $J=J_{\mathsf{A}}+\sigma_n\mathsf{I}$, with $J_{\mathsf{A}}$ the Jacobian of the linear game and
\begin{equation}
\sigma_n(a)=\frac{1}{n}\varphi'\!\left(\frac1n\right)=\frac{an(n-2)-1}{n(1+an)^2},
\label{eq:sigma}
\end{equation}
where we used $\varphi'(u)=[a(1-2u)-u^2]/(u+a)^2$. The self-term shifts every eigenvalue by the same real amount. The shift is negative (toward stability) iff $a<a_c(n)$ and positive (toward instability) iff $a>a_c(n)$, where
\begin{equation}
a_c(n)=\frac{1}{n(n-2)}\quad(n\ge3),
\label{eq:ac}
\end{equation}
and it is always negative for $n=2$. Whether it actually destabilizes the barycenter depends on the real parts of the eigenvalues of $J_{\mathsf{A}}$. These hypotheses are restrictive, since for a general payoff matrix the Jacobian contains additional terms.

\begin{figure*}[t]
\includegraphics[width=\textwidth]{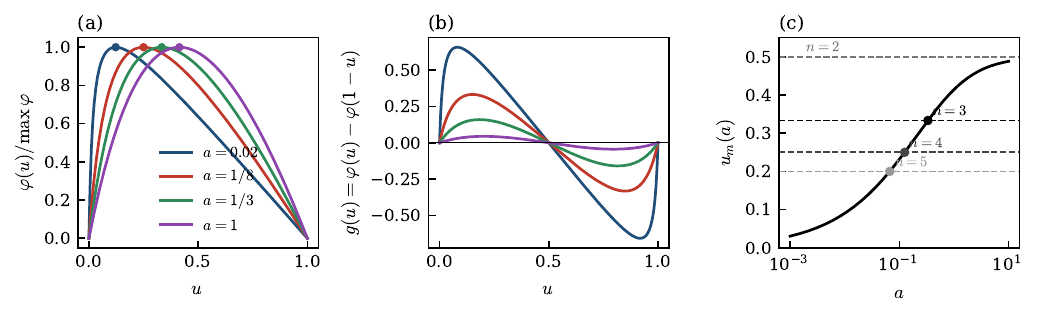}
\caption{(a) The mate-finding Allee function $\varphi(u)=u(1-u)/(u+a)$, normalized by its maximum, for several values of $a$. Dots mark the maximum at $u_m(a)$, Eq.~\eqref{eq:um}. (b) The two-strategy self-term $g(u)=\varphi(u)-\varphi(1-u)$, Eq.~\eqref{eq:g}. It is positive for $u<1/2$, i.e., it favors the rare strategy for every $a$. (c) $u_m(a)$ (solid line). Its intersections with $u=1/n$ (dashed lines) give the critical values $a_c(n)=1/[n(n-2)]$ of Eq.~\eqref{eq:ac}. The curve never reaches the line $u=1/2$.}
\label{fig:phi}
\end{figure*}

The threshold \eqref{eq:ac} is equivalent to the condition $u_m(a)=1/n$ [Fig.~\ref{fig:phi}(c)]. The self-term pushes the symmetric state toward instability precisely when the common frequency $1/n$ lies on the Allee branch of $\varphi$. For $n=2$ we have $u_m<1/2$ for every $a$, and the barycenter always lies on the crowding branch. It is therefore a \emph{strong} Allee effect (large $a$) that can destabilize coexistence. With a weak Allee effect, crowding dominates $\varphi$, with $\varphi(u)\simeq1-u$ for $u\gg a$, and acts as a negative frequency dependence. We obtain $a_c=1/3$ for $n=3$, $a_c=1/8$ for $n=4$, and $a_c\simeq n^{-2}$ for large $n$.

\section{Two strategies}
\label{sec:two}

\subsection{Rarity advantage}

For $n=2$ we write $u=x_1$. With $\alpha=A_{11}-A_{21}$ and $\beta=A_{22}-A_{12}$, Eq.~\eqref{eq:rep} reduces to
\begin{equation}
\dot u=u(1-u)H(u),\qquad H(u)=g(u)+(\alpha+\beta)u-\beta,
\label{eq:n2}
\end{equation}
where the contribution of the self-term is
\begin{equation}
\begin{split}
g(u)&=\varphi(u)-\varphi(1-u)\\
&=(1-2u)\left[1-\frac{a(1+a)}{(u+a)(1+a-u)}\right].
\end{split}
\label{eq:g}
\end{equation}
Since $(u+a)(1+a-u)\ge a(1+a)$, with equality only at $u\in\{0,1\}$, we have $g>0$ on $(0,\tfrac12)$ and $g<0$ on $(\tfrac12,1)$ [Fig.~\ref{fig:phi}(b)]. This result has a simple interpretation. If we write $\varphi(u)=(1-u)\,u/(u+a)$, we see two opposite effects acting on the rare type. On the one hand, it suffers from mate limitation, $u/(u+a)$. On the other hand, it experiences a weaker crowding, $1-u$. For two types that share the numerator $u(1-u)$, the second effect always wins. \emph{At constant density and for two strategies, the type-specific crowding thus dominates the mate-finding Allee effect, and $\varphi$ produces a negative frequency dependence}, the same stabilizing mechanism of anticoordination games, superimposed on the game. In fact, for $a\to0$ we have $\varphi\to1-u$ and $g\to1-2u$, so the rarity advantage is already present without any mate limitation, which only weakens it (see $G(a)$ below). We also have $g(0)=g(\tfrac12)=g(1)=0$ and
\begin{equation}
g\!\left(\tfrac12+z\right)=-\frac{2z}{(1+2a)^2}+\frac{32a(1+a)}{(1+2a)^4}z^3+O(z^5).
\label{eq:gexp}
\end{equation}
The maximum $G(a)=\max_u g(u)$ decreases monotonically from $G\to1$ as $a\to0$.

By Eq.~\eqref{eq:vertex}, the eigenvalues at $u=0$ and $u=1$ are $H(0)=-\beta$ and $-H(1)=-\alpha$, as in the linear game. The self-term only reshapes the interior.

\subsection{Interior equilibria}

The function $g$ is strictly concave on $(0,\tfrac12)$ and strictly convex on $(\tfrac12,1)$. To see it, we write $z=u-\tfrac12$, $c=\tfrac12+a$ and $k=a(1+a)=c^2-\tfrac14$, Eq.~\eqref{eq:g} reads $g=-2z+k[(c-z)^{-1}-(c+z)^{-1}]$. Then $g''=2k[(c-z)^{-3}-(c+z)^{-3}]$, which is negative for $z<0$ and positive for $z>0$.

Since $H$ differs from $g$ by a linear function, $H$ inherits these properties and has at most two zeros on each half-interval. As stability alternates along the line, the signs of $H(0)=-\beta$ and $H(1)=\alpha$ fix the parity of the number of interior equilibria. The complete classification follows from these properties (Appendix~\ref{app:count}).

In dominance games ($\alpha<0<\beta$), as the prisoner's dilemma with $u$ the frequency of cooperators, there are zero or two interior equilibria, both in $(0,\tfrac12)$. In the second case the system is bistable, and full defection coexists with a stable mixed state $u_2^*<\tfrac12$, separated from it by an unstable threshold $u_1^*$. Cooperators can persist only as a minority.

In coordination games ($\alpha,\beta>0$), as the stag hunt, there are one or three interior equilibria. With three, the middle one is stable and the dynamics becomes \emph{tristable}, with attractors at $u=0$, $u=1$, and in the interior.

In anticoordination games ($\alpha,\beta<0$), as hawk-dove, there is exactly one interior equilibrium, which attracts all of $(0,1)$. With respect to the linear game, it moves toward $u=1/2$.

The boundaries between these regimes are saddle-node points, $H(u)=H'(u)=0$. Since $H$ is linear in $(\alpha,\beta)$, the bifurcation set is the envelope of the family of lines $H(u;\alpha,\beta)=0$, which we can parametrize by the location $u$ of the double root:
\begin{equation}
\beta_{\rm sn}(u)=g(u)-ug'(u),\qquad \alpha_{\rm sn}(u)=-g(u)-(1-u)g'(u),
\label{eq:env}
\end{equation}
with $u\in(0,1)$. The envelope has a cusp at $\alpha=\beta=1/(1+2a)^2$ ($u=1/2$) [Fig.~\ref{fig:n2}(a)]. The two branches meet at the cusp and together bound the region with three equilibria inside the coordination quadrant. The branch $u\in(0,\tfrac12)$ also bounds the region with two equilibria inside the dominance quadrant $\alpha<0<\beta$, and the branch $u\in(\tfrac12,1)$ the corresponding region of the mirror quadrant $\beta<0<\alpha$.

\subsection{Two special cases}

\begin{figure*}[t]
\includegraphics[width=\textwidth]{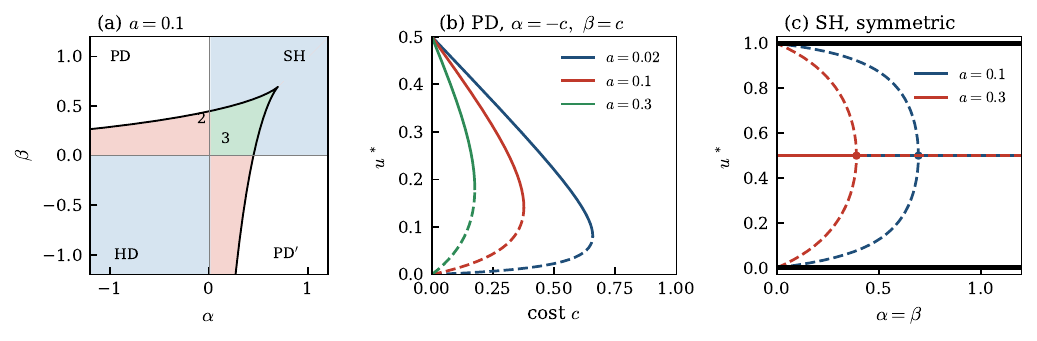}
\caption{Two-strategy games. (a) Number of interior equilibria in the $(\alpha,\beta)$ plane for $a=0.1$: one (blue), two (red), three (green), none (white). The solid curve is the saddle-node envelope \eqref{eq:env}, with a cusp at $\alpha=\beta=(1+2a)^{-2}$. The quadrant labels refer to the linear game: dominance of strategy 2 (PD), dominance of strategy 1 (PD$'$), coordination (SH), and anticoordination (HD). (b) Interior equilibria of the prisoner's dilemma with equal gains from switching as a function of the cost $c$. Solid lines indicate stable and dashed lines unstable equilibria; the saddle-node bifurcation occurs at $c=G(a)$. (c) Symmetric stag hunt: equilibria as a function of $\alpha=\beta$ for $a=0.1$ and $a=0.3$. Thick black lines mark the stable pure states. The center is stable for $\alpha<(1+2a)^{-2}$ and loses stability in a subcritical pitchfork.}
\label{fig:n2}
\end{figure*}

In the prisoner's dilemma with equal gains from switching, $\alpha=-c$ and $\beta=c$, we have $H=g(u)-c$, and the interior equilibria satisfy $g(u^*)=c$. They exist iff $c<G(a)$, and they appear through a saddle-node bifurcation at $c=G(a)$ [Fig.~\ref{fig:n2}(b)]. For example, $G=0.658$, $0.376$, and $0.176$ for $a=0.02$, $0.1$, and $0.3$.

In the symmetric stag hunt, $\alpha=\beta$, the state $u=\tfrac12$ is always an equilibrium and, by Eq.~\eqref{eq:gexp},
\begin{equation}
H\!\left(\tfrac12+z\right)=2\left[\alpha-\frac{1}{(1+2a)^2}\right]z+\frac{32a(1+a)}{(1+2a)^4}z^3+O(z^5).
\end{equation}
The center is stable for $\alpha<\alpha_c=(1+2a)^{-2}$. Since the cubic coefficient is positive, the two additional equilibria that exist for $\alpha<\alpha_c$ are unstable, and they merge with the center at $\alpha_c$ in a subcritical pitchfork [Fig.~\ref{fig:n2}(c)]. A weak coordination ($\alpha<\alpha_c$) therefore does not turn bistability into global coexistence. Instead, it produces \emph{tristability}, with an interior basin bounded by two thresholds that move toward the pure states as $\alpha\to0$. For $\alpha\neq\beta$ the pitchfork unfolds into the saddle-node curves \eqref{eq:env}.

\section{Rock-paper-scissors}
\label{sec:rps}

\subsection{Hopf curve}
\label{sec:hopf}

We take
\begin{equation}
\mathsf{A}=\begin{pmatrix}0&-l&w\\ w&0&-l\\ -l&w&0\end{pmatrix},\qquad w,l>0,
\label{eq:Arps}
\end{equation}
with win payoff $w$, loss $l$, asymmetry $\delta=w-l$, and normalization $w+l=2$ unless we state otherwise. In the linear game the barycenter is globally stable for $\delta>0$, while for $\delta<0$ the heteroclinic cycle $\mathbf e_1\to\mathbf e_2\to\mathbf e_3\to\mathbf e_1$ on the boundary attracts the orbits. The zero-sum case $\delta=0$ is conservative, with first integral $V=x_1x_2x_3$~\cite{HofbauerSigmund1998,Zeeman1980}. The heteroclinic cycle persists as long as no equilibria appear on the edges. Since the velocity along an edge is $u(1-u)[w(1-u)+lu+g(u)]$, the condition $\min(w,l)>G(a)$ guarantees it. By Eq.~\eqref{eq:vertex}, the linear stability of the cycle does not change: it attracts iff $(l/w)^3>1$, i.e., $\delta<0$~\cite{HofbauerSigmund1998,KrupaMelbourne1995}. By Eq.~\eqref{eq:sigma}, the eigenvalues at the barycenter are
\begin{equation}
\lambda_\pm=\frac{3a-1}{3(1+3a)^2}-\frac{\delta}{6}\pm i\,\frac{\sqrt3\,(w+l)}{6},
\label{eq:eig}
\end{equation}
and the center loses stability through a Hopf bifurcation on the curve
\begin{equation}
\delta_H(a)=\frac{2(3a-1)}{(1+3a)^2},
\label{eq:hopf}
\end{equation}
which does not depend on $w+l$. The curve crosses $\delta=0$ at $a_c=1/3$ and reaches its maximum $\delta_H=1/4$ at $a=1$. A strong Allee effect can therefore destabilize the RPS equilibrium even when wins exceed losses, $0<\delta<1/4$, and a weak one can stabilize it even when losses exceed wins.

Except at a single point, the Hopf bifurcation is non-degenerate. On the curve \eqref{eq:hopf} we computed the first Lyapunov coefficient $\ell_1$~\cite{Kuznetsov2004,GuckenheimerHolmes1983} in closed form (Appendix~\ref{app:l1}),
\begin{equation}
\ell_1(a)=\frac{27\sqrt3}{w+l}\,\frac{a(1+a)(2-3a)}{2(1+3a)^4},
\label{eq:l1}
\end{equation}
where the prefactor depends on the normalization of the critical eigenvector but the sign does not. The coefficient changes sign at a Bautin (generalized Hopf) point $B$,
\begin{equation}
a_B=\frac23,\qquad \delta_B=\delta_H(a_B)=\frac29,
\label{eq:bautin}
\end{equation}
for every payoff scale $w+l$. The Hopf bifurcation is subcritical ($\ell_1>0$) for $a<a_B$ and supercritical ($\ell_1<0$) for $a>a_B$ [Fig.~\ref{fig:diagram}(a)].

\subsection{Zero-sum game}

For $\delta=0$ the function $V=x_1x_2x_3$ is no longer a constant of motion. Using $\sum_i(\mathsf{A}\x)_i=0$ and $\x^{\!\top}\mathsf A\x=0$, we obtain the \emph{exact} identity
\begin{equation}
\begin{split}
\frac{\dd\ln V}{\dd t}&=\sum_i\varphi(x_i)-3\sum_ix_i\varphi(x_i)\\
&=-3\sum_i\left(x_i-\tfrac13\right)\varphi(x_i).
\end{split}
\label{eq:lnV}
\end{equation}
Near the center, with $\x=\bar\x+\mathbf v$, Eq.~\eqref{eq:lnV} gives $\dd\ln V/\dd t\simeq-3\varphi'(\tfrac13)|\mathbf v|^2$, where $\varphi'(\tfrac13)=(3a-1)/(1+3a)^2$. For $a<1/3$, $V$ grows toward its maximum at $\bar\x$ and the center is stable, while for $a>1/3$ the center is unstable. At $a=1/3$ the linear term vanishes, but the first Lyapunov coefficient does not, since Eq.~\eqref{eq:l1} gives $\ell_1>0$ there (Sec.~\ref{sec:hopf}). The orbits are therefore not closed, and $a=1/3$ is a true Hopf point and not a center.

The identity \eqref{eq:lnV} is a particular case of a more general one. If we add an arbitrary self-term to the fitness of zero-sum RPS, $f_i=(\mathsf A\x)_i+s_i(\x)$, the same steps give $\dd\ln V/\dd t=\sum_i(1-3x_i)\,s_i(\x)$, so the self-term pushes the orbits inward or outward according to the relative contribution of the types below and above $1/3$. For $s_i=\varphi(x_i)$ the slope of $\varphi$ at $1/3$ decides the balance. A term that instead suppresses the fitness of rare types behaves differently. If a viability factor $\chi(x_i)\in[0,1]$ multiplies a positive fitness $c+(\mathsf A\x)_i$, equals one above some frequency smaller than $1/3$, and vanishes below a lower threshold, then $s_i=-[1-\chi(x_i)][c+(\mathsf A\x)_i]$ (the constant $c$ cancels in the identity). This self-term is nonpositive and acts only on types with $x_i<1/3$, so it always drives the orbits toward the boundary.

\begin{figure*}[t]
\includegraphics[width=\textwidth]{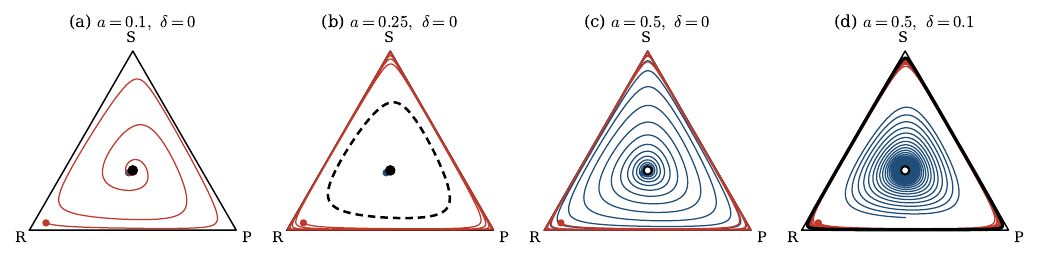}
\caption{Phase portraits of RPS on $\Delta^2$, with vertices R, P, and S. The filled (open) circle marks the stable (unstable) barycenter. (a) $a=0.1<a_h$, $\delta=0$: the center attracts all interior orbits. (b) $a=0.25$, $\delta=0$: bistability between the center and the heteroclinic cycle, with an unstable limit cycle (dashed) separating the basins. (c) $a=0.5$, $\delta=0$: the center is unstable and the orbits accumulate on the boundary. (d) $a=0.5$, $\delta=0.1$: a stable limit cycle (thick line) attracts the orbits from inside and outside. In the linear game with $\delta=0.1$ the center would be globally stable.}
\label{fig:rps}
\end{figure*}

Equation~\eqref{eq:lnV} also determines the stability of the boundary. Close to the heteroclinic cycle, the trajectories follow the edges and spend a logarithmically long time near the vertices. There $\dd\ln V/\dd t$ is proportional to the distance to the vertex, and the vertex passages do not contribute in the limit. The net change of $\ln V$ per revolution is thus three times the integral of Eq.~\eqref{eq:lnV} along one edge (Appendix~\ref{app:het}):
\begin{equation}
\mathcal I(a)=\int_0^1\frac{1-a-6u(1-u)}{(u+a)(1+a-u)\,[w+g(u)]}\,\dd u\qquad(w=l).
\label{eq:Ia}
\end{equation}
Here $u$ is the frequency of the invader along the edge, which moves with velocity $\dot u=u(1-u)[w+g(u)]$. The heteroclinic cycle attracts the orbits if $\mathcal I<0$ and repels them if $\mathcal I>0$. Numerically, the integral $\mathcal I(a)$ changes sign only once in the range we explored, at
\begin{equation}
a_h\simeq0.1336\quad(w=l=1),
\end{equation}
and depends only weakly on the payoff scale: $a_h\simeq0.1324$ for $w=l=2$ and $0.1321$ for $w=l=4$ [Fig.~\ref{fig:diagram}(c)]. At $\delta=0$ we then find, numerically, three regimes. For $a<a_h$ the center attracts all of $\mathrm{int}\,\Delta^2$. For $a_h<a<1/3$ the center and the heteroclinic cycle are both attracting. Since the annulus between them contains no equilibria and repels the orbits in reverse time, the Poincar\'e-Bendixson theorem guarantees at least one periodic orbit in it, and numerically we find a single \emph{unstable limit cycle}, which separates the two basins. This cycle appears in the subcritical Hopf bifurcation at $a=1/3$ with an amplitude $\propto(1/3-a)^{1/2}$, grows as $a$ decreases, and merges with the heteroclinic cycle at $a=a_h$ [Figs.~\ref{fig:rps}(b) and \ref{fig:diagram}(b)]. For $a>1/3$ all interior orbits except the center converge to the boundary.

\subsection{Bifurcation diagram}

\begin{figure*}[t]
\includegraphics[width=\textwidth]{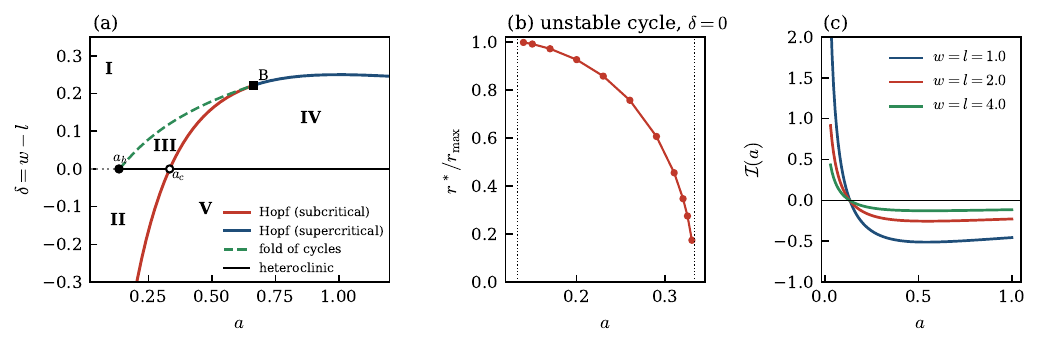}
\caption{(a) Bifurcation diagram of RPS in the $(a,\delta)$ plane ($w+l=2$). The Hopf curve \eqref{eq:hopf} is subcritical (red) below the Bautin point $B$ and supercritical (blue) above it. The dashed curve is the fold of limit cycles, $\delta_F(a)$. The horizontal line $\delta=0$ separates attracting ($\delta<0$) from repelling ($\delta>0$) heteroclinic cycles; on $\delta=0$ the cycle attracts the orbits for $a>a_h$ (solid) and repels them for $a<a_h$ (dotted). We describe regions I--V in the text. (b) Amplitude of the unstable limit cycle at $\delta=0$, measured as the distance $r^*$ from the center along the Poincar\'e section and normalized by its maximum $r_{\max}=\sqrt2/3$. (c) The heteroclinic stability integral $\mathcal I(a)$, Eq.~\eqref{eq:Ia}, for several payoff scales; its zero gives $a_h$.}
\label{fig:diagram}
\end{figure*}

For $\delta\neq0$, the sign of $\delta$ determines the stability of the boundary at leading order [Eq.~\eqref{eq:vertex}]. If we combine this fact with Eq.~\eqref{eq:hopf}, the Bautin point, and $a_h$, we can divide the $(a,\delta)$ plane into five regions [Fig.~\ref{fig:diagram}(a)]. We confirmed this partition by counting the fixed points of the return map on a grid of $47\times53$ parameter values. In region I, with $\delta>0$ above the Hopf and fold curves, the center is globally stable. In region II, $\delta_H<\delta<0$, the center and the heteroclinic cycle are both attracting. As in the regime $a_h<a<1/3$ at $\delta=0$, at least one periodic orbit separates their basins, and numerically we find a single unstable limit cycle. For $a<a_h$ this cycle lies exponentially close to the boundary as $\delta\to0^-$. In region III, $0<\delta<\delta_F(a)$ and $\delta>\delta_H$, an unstable cycle and an outer \emph{stable} cycle surround the stable center. Coexistence at equilibrium and sustained oscillations of large amplitude are then both attracting, with hysteresis. In region IV, $0<\delta<\delta_H$ (only for $a>1/3$), the center is unstable and the boundary repels the orbits. By the Poincar\'e-Bendixson theorem applied to the annulus between them, at least one stable limit cycle exists. In all the cases that we sampled we find a single one, which attracts all interior orbits [Fig.~\ref{fig:rps}(d)], but we do not prove its uniqueness. In region V, $\delta<\min(0,\delta_H)$, the orbits converge to the heteroclinic cycle.

We obtained by bisection the fold-of-cycles curve $\delta_F(a)$ (for $w+l=2$), where the stable and unstable cycles of region III collide. The curve starts at $(a_h,0)$, takes the values $\delta_F\simeq0.05$ at $a=0.2$ and $\delta_F\simeq0.11$ at $a=0.3$, and ends tangentially on the Hopf curve at the Bautin point. At $B$ the second Lyapunov coefficient is negative (Appendix~\ref{app:l1}), so the Bautin point is nondegenerate, and the fold of cycles emerges on the subcritical side, as we observe. Region III is the codimension-one unfolding of both the degenerate heteroclinic point $(a_h,0)$ and the Bautin point.

Stable isolated periodic orbits, which cannot exist in the three-strategy linear replicator dynamics~\cite{Zeeman1980}, therefore occur in the extended regions III and IV, with an existence guaranteed analytically in region IV and supported numerically in region III. They arise from a single demographic term, without mutations~\cite{Mobilia2010,ToupoStrogatz2015}, higher-order interactions~\cite{GriffinWu2023}, or noise. It is worth mentioning that Griffin and Wu~\cite{GriffinWu2023} found a subcritical Hopf bifurcation and an unstable cycle in RPS with triadic interactions. The Allee term reproduces this scenario in region II, and it adds the supercritical branch, the Bautin point, and the stable cycles.

\section{Free total density}
\label{sec:freeN}

The results above rely on the fact that we fix the total density at $N=K$. Without regulation, $D\equiv0$, and in terms of $n_i=N_i/K$ Eq.~\eqref{eq:demo} reads
\begin{equation}
\dot n_i=n_i\big[\varphi(n_i)-m+(\mathsf A\x)_i\big],\qquad \x=\mathbf n/\nu,\quad\nu=\textstyle\sum_kn_k.
\label{eq:free}
\end{equation}
The frequencies then obey Eq.~\eqref{eq:rep} with $\varphi(\nu x_i)$ in place of $\varphi(x_i)$, coupled to $\dot\nu=\nu[\sum_kx_k\varphi(\nu x_k)-m+\x^{\!\top}\mathsf A\x]$. The mortality is no longer irrelevant. A single type now obeys exactly the one-dimensional Allee model \eqref{eq:allee1d}, with fixed points $n=0$ (stable), $n_A=u_A$ (the Allee threshold), and $n_C=u_C$ (stable), given by Eq.~\eqref{eq:uAC}.

At a symmetric RPS state $n_i=n^*$, with $\varphi(n^*)=m-\delta/3$, the radial (ecological) and tangent (evolutionary) modes decouple, with eigenvalues
\begin{equation}
\lambda_\nu=n^*\varphi'(n^*),\qquad \mathrm{Re}\,\lambda_\pm=n^*\varphi'(n^*)-\frac\delta6.
\end{equation}
Ecological stability, $\lambda_\nu<0$, requires that $n^*$ lies on the crowding branch, $n^*>u_m(a)$, and there the self-term \emph{always stabilizes} the frequencies. The Allee-induced destabilization of Sec.~\ref{sec:rps} occurs only when the types lie on the Allee branch of $\varphi$, a situation that is ecologically unstable in an unregulated population. The destabilization of the center, and with it regime IV, is specific to regulated populations, such as chemostat or turbidostat cultures, or populations of fixed size. We did not integrate Eq.~\eqref{eq:free} in regime III, so we do not know whether its large-amplitude stable cycles survive when the density is free.

\begin{figure*}[t]
\includegraphics[width=\textwidth]{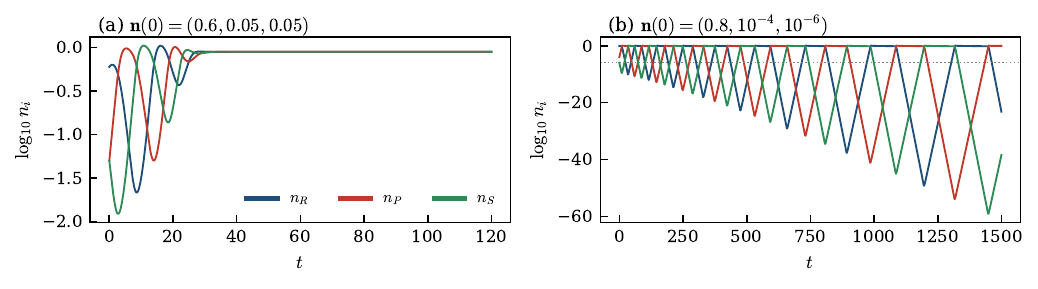}
\caption{Free total density, Eq.~\eqref{eq:free}, for RPS with $a=0.5$, $m=0.1$, and $\delta=0.1<2m$. (a) An initial condition near the barycenter converges to coexistence at $n_i\simeq0.896$. (b) An initial condition near a vertex converges to the heteroclinic cycle, and the minimum densities decrease without bound.}
\label{fig:free}
\end{figure*}

The Allee mortality acts instead at the boundary. At a vertex the density of the resident is $n_C$, and the invaders grow at rate $A_{ji}-m$ instead of $A_{ji}-A_{ii}$. For $w>m$ each vertex of RPS is a saddle with one expanding direction. The phase space is now three dimensional, and the criterion of Sec.~\ref{sec:rps} does not apply directly. However, the coordinate planes $n_k=0$ remain invariant and each connection lies in one of them, so that the vertices form a simple robust heteroclinic cycle in $\mathbb R^3$, of the same type as that of the May-Leonard system~\cite{MayLeonard1975,KrupaMelbourne1995}. At each vertex the eigenvalues split into a radial one, along the axis of the resident, which is negative, a contracting one, $-(l+m)$, and an expanding one, $w-m$, with no transverse eigenvalues. By the criterion of Krupa and Melbourne~\cite{KrupaMelbourne1995}, the radial eigenvalue enters only through its sign, and the cycle attracts the orbits iff $(l+m)^3>(w-m)^3$, i.e., for $\delta<2m$. We observed the existence of the connections numerically but did not prove it. For $w<m$ the vertices become stable. We integrated Eq.~\eqref{eq:free} for $a=0.5$ and $m=0.1$ ($n_C=0.8405$ at $\delta=0$). The initial conditions near the barycenter converge to coexistence for all values of $\delta$ that we explored. The initial conditions sufficiently close to the boundary converge to the heteroclinic cycle for $\delta<2m=0.2$, with a basin that shrinks as $\delta\to0.2^-$, and to coexistence for $\delta>0.2$ (Fig.~\ref{fig:free}). Along the attracting heteroclinic cycle the minimum densities decrease without bound and reach $\log_{10}n_i<-50$ within $t\sim10^3$, which in any finite population means the loss of that type.

\section{Shared crowding}
\label{sec:crowd}

The rarity advantage of Sec.~\ref{sec:two} comes from the type-specific crowding, and it relies on the choice $N=K$. To test both assumptions, we assume that the crowding depends on a mixture of the density of the type and the total density, and we fix the total density at $N=\rho K$. The per-capita birth rate of type $i$ becomes
\begin{equation}
\frac{N_i}{N_i+\theta}\left[1-\frac{\kappa N_i+(1-\kappa)N}{K}\right],\qquad 0\le\kappa\le1,
\label{eq:kappa}
\end{equation}
where $\kappa=1$ is the type-specific crowding of Eq.~\eqref{eq:demo} and $\kappa=0$ is the crowding by a resource that all types share. For $\rho=\kappa=1$ we recover the model of Sec.~\ref{sec:model}. With $N_i=\rho x_iK$, the derivation of Sec.~\ref{sec:model} gives Eq.~\eqref{eq:rep} with $f_i=\Phi(x_i)+(\mathsf A\x)_i$, where
\begin{equation}
\begin{split}
\Phi(x)&=\frac{\rho x}{\rho x+a}\big[1-\rho(1-\kappa)-\rho\kappa x\big]\\
&=(1-\rho)\,\frac{x}{x+\tilde a}+\kappa\rho\,\frac{x(1-x)}{x+\tilde a},\qquad\tilde a=\frac a\rho.
\end{split}
\label{eq:Phi}
\end{equation}
The first term is pure mate finding, and it increases with $x$. The second term is the function $\varphi$ of Eq.~\eqref{eq:allee1d} with $a$ replaced by $\tilde a$ and weighted by $\kappa\rho$. As in Sec.~\ref{sec:model}, the reduction is exact once we fix the total density.

Two changes follow, and both follow from the arguments of Sec.~\ref{sec:general} with $\varphi$ replaced by $\Phi$. First, for $\rho<1$ we have $\Phi(1)=\varphi_1=\rho(1-\rho)/(\rho+a)>0$, and the invasion exponents at a vertex become $\lambda_{j|i}=A_{ji}-A_{ii}-\varphi_1$. The growth rate of a rare invader decreases by $\varphi_1$, since it does not find mates. This is a genuine Allee effect, which makes both pure states of a two-strategy game stable whenever $\alpha,\beta>-\varphi_1$, and which makes the heteroclinic cycle of RPS attracting for $\delta<2\varphi_1$ (with $w>\varphi_1$), the counterpart of the condition $\delta<2m$ of Sec.~\ref{sec:freeN}. A viability threshold that multiplies the fitness protects the pure states in every game. Here the protection appears only for $\rho<1$, and its strength is bounded by $\varphi_1$.

\begin{figure*}[t]
\includegraphics[width=\textwidth]{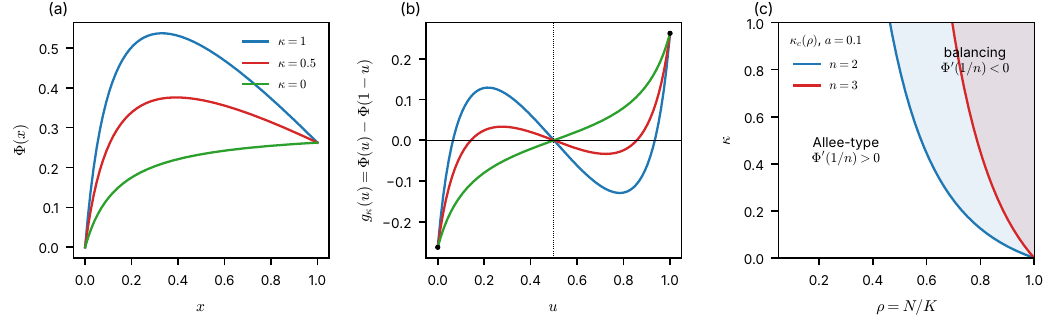}
\caption{Interpolation between type-specific ($\kappa=1$) and shared ($\kappa=0$) crowding at a fixed density $N=\rho K$. (a) The self-term $\Phi(x)$, Eq.~\eqref{eq:Phi}, for $\rho=0.7$, $a=0.1$, and $\kappa=1$, $0.5$, and $0$. (b) The two-strategy self-term $g_\kappa(u)=\Phi(u)-\Phi(1-u)$, Eq.~\eqref{eq:gk}, for the same parameters. The values at the ends are $g_\kappa(0)=-\varphi_1$ and $g_\kappa(1)=\varphi_1$ (dots). (c) Critical weight $\kappa_c(\rho)$ of the type-specific crowding for $a=0.1$, for $n=2$ and $n=3$. Above each curve (shaded) the self-term is balancing at the symmetric state, $\Phi'(1/n)<0$, and below it the self-term is Allee-type, $\Phi'(1/n)>0$.}
\label{fig:crowd}
\end{figure*}

Second, $\kappa$ controls the sign of the effect at the symmetric state, through $\Phi'(x)=\{(1-\rho)\tilde a+\kappa\rho[\tilde a(1-2x)-x^2]\}/(x+\tilde a)^2$. For $n=2$ the self-term reads
\begin{equation}
\begin{split}
g_\kappa(u)&=\Phi(u)-\Phi(1-u)\\
&=(1-2u)\,\frac{\kappa\rho\,u(1-u)-(1-\rho)\tilde a}{(u+\tilde a)(1+\tilde a-u)},
\end{split}
\label{eq:gk}
\end{equation}
and it is balancing at $u=\tfrac12$ iff $\kappa>\kappa_c(\rho)=4(1-\rho)a/\rho^2$ [Fig.~\ref{fig:crowd}(c)]. In that case the rarity advantage survives in the interior but not close to the pure states, where the rare type suffers this cost. For $n\ge3$, a balancing self-term requires $\tilde a<a_c(n)$ and $\kappa>\kappa_c(n)=(1-\rho)\tilde a\big/\{\rho[1/n^2-\tilde a(1-2/n)]\}$, so that a regulated density below $K$ shifts this condition from $a<a_c(n)$ to $a<\rho\,a_c(n)$. In RPS the Hopf curve becomes
\begin{equation}
\delta_H=2\Phi'\!\left(\tfrac13\right)=\frac{2\left[(1-\rho)\tilde a+\kappa\rho\,(\tilde a/3-1/9)\right]}{(1/3+\tilde a)^2},
\label{eq:hopfG}
\end{equation}
which reduces to Eq.~\eqref{eq:hopf} for $\rho=\kappa=1$.

In the shared limit $\kappa=0$, $g_0(u)$ is negative for $u<\tfrac12$ [Fig.~\ref{fig:crowd}(b)]. The self-term then penalizes the rare type and acts as a positive frequency dependence. In the prisoner's dilemma a cooperator minority can no longer persist, and with equal gains from switching, $\alpha=-c$ and $\beta=c$, both pure states are stable for $c<\varphi_1$, separated by a critical mass of cooperators. For intermediate values of $\kappa$ both mechanisms coexist. With $\rho=0.7$, $a=0.1$, $\kappa=1$, and $c=0.05$, for instance, the prisoner's dilemma becomes tristable, with defection, a cooperator minority at $u\simeq0.420$, and full cooperation as attractors. In RPS with $\kappa=0$, $\rho=\tfrac12$, and $a=0.3$, where $\delta_H\simeq0.689$ and $2\varphi_1=0.625$, we find numerically the sequence of attractors of regions V, IV, and I of Sec.~\ref{sec:rps} as $\delta$ grows, with the threshold of the boundary moved from $\delta=0$ to $\delta=2\varphi_1$.

\section{Discussion}

We have analyzed a simple model in which each strategy of an evolutionary game carries its own mate-finding Allee term, and we have shown that the effects of this term depend strongly on the number of strategies and on which branch of $\varphi$ the population visits. For two strategies the term acts as a rarity advantage, which allows a minority of cooperators to persist in the prisoner's dilemma and makes coordination games tristable. For three or more strategies the sign of its effect at the symmetric state depends on the position of $1/n$ with respect to the maximum of $\varphi$. In RPS, this mechanism produces a Hopf curve and a Bautin point in closed form, and isolated stable and unstable limit cycles. When the total density is free, the same term stabilizes coexistence and widens the range of payoffs for which the heteroclinic cycle attracts the orbits.

There is however a caveat: the strategy-specific self-term presupposes that each type has its own mating pool and its own crowding, i.e., some degree of niche or reproductive separation. This assumption is natural for competing species or strains, as in cyclic microbial systems~\cite{Kerr2002,Szolnoki2014}, and less natural for behavioral strategies within a single interbreeding population, as in the lizard morphs of Ref.~\cite{Sinervo1996}. The sign of the effect depends on this assumption. If the crowding depends on the total density instead, the mate-finding term increases with the frequency and penalizes the rare type, and a genuine Allee effect replaces the rarity advantage (Sec.~\ref{sec:crowd}). For the same reason, the coexistence of cooperators in the prisoner's dilemma (Sec.~\ref{sec:two}) is not a mechanism for the evolution of cooperation in the sense of Ref.~\cite{Nowak2006}. It results from the rarity advantage that type-specific crowding gives to whichever strategy is rare. Its signatures are nevertheless testable: bistability with a defection-only state, and a coexistence frequency below $1/2$. In the same way, the Allee-driven destabilization of the center in RPS requires a regulated total density (Sec.~\ref{sec:freeN}). Finally, the direction of the effect depends on how the term enters the fitness, added to it as here, or multiplying it as a viability threshold would (Sec.~\ref{sec:rps}).

Within these limits, the model has some interesting features. Besides the payoff scale and, in Sec.~\ref{sec:freeN}, the mortality $m$, it has a single interpretable parameter, $a$, and it leads to the closed-form results \eqref{eq:sigma}, \eqref{eq:ac}, \eqref{eq:env}, \eqref{eq:hopf}, \eqref{eq:l1}, and \eqref{eq:Ia}. It also offers a transparent route to isolated limit cycles in three-strategy games. Natural extensions include stochastic versions of the constant-density model, where the unstable cycle of region II determines the time to fixation, spatially extended RPS~\cite{Reichenbach2007,Szolnoki2014}, games on networks~\cite{Abramson2001}, and larger cyclic games, where Eq.~\eqref{eq:sigma} predicts destabilization already for $a>1/[n(n-2)]$.



\appendix
\section{Classification for $n=2$}
\label{app:count}

The function $H=g+\ell$, with $\ell(u)=(\alpha+\beta)u-\beta$ linear, is concave on $[0,\tfrac12]$ and convex on $[\tfrac12,1]$, and it has at most two zeros on each half. The interior equilibria are the zeros of $H$, and their stability alternates. Therefore, the number of interior zeros, counted with multiplicity, is even if $H(0)H(1)>0$ and odd otherwise.

In dominance games, $\alpha<0<\beta$, we have $\ell<0$ on $[0,1]$. On $[\tfrac12,1]$ we have $g\le0$, and therefore $H<0$. On $[0,\tfrac12]$, $H$ is concave with $H(0)=-\beta<0$ and $H(\tfrac12)=\ell(\tfrac12)<0$, and it has zero or two zeros.

In anticoordination games, $\alpha,\beta<0$, we have $H(0)>0>H(1)$. If $H(\tfrac12)\ge0$, concavity forbids zeros on $(0,\tfrac12)$, since $H$ lies above its chord. Convexity, together with the odd parity, allows exactly one zero on $[\tfrac12,1)$. The case $H(\tfrac12)<0$ is symmetric. The unique equilibrium is stable because $H$ changes sign from positive to negative.

In coordination games, $\alpha,\beta>0$, we have $H(0)<0<H(1)$. If $H(\tfrac12)>0$ there is exactly one zero in $(0,\tfrac12)$ and zero or two in $(\tfrac12,1)$, which gives one or three in total. With three zeros, stability alternates as unstable, stable, unstable. The case $H(\tfrac12)\le0$ is symmetric.

The double-root conditions $H=H'=0$ give $\alpha+\beta=-g'(u)$ and $\beta=g(u)-ug'(u)$, which yield Eq.~\eqref{eq:env}.

\section{Heteroclinic stability}
\label{app:het}

Let $Q(\x)=\dd\ln V/\dd t$, given by Eq.~\eqref{eq:lnV}. We consider a Poincar\'e section transverse to the incoming edge near a vertex, at a fixed distance $\epsilon_0$ from the vertex. On this section, $\ln V=\ln\epsilon+\ln\epsilon_0+O(1)$, where $\epsilon$ is the distance to the boundary. The change of $\ln\epsilon$ after one revolution therefore equals $\oint Q\,\dd t$. Near a vertex, $Q\simeq(1-a)(x_j+x_k)/[a(1+a)]$, linear in the small coordinates. The contribution of a vertex passage is thus $O(\epsilon_0)$ and vanishes as $\epsilon_0\to0$. Along the edge $x_3=0$, with $x_2=u$, a direct calculation gives
\begin{equation}
Q=\frac{u(1-u)[1-a-6u(1-u)]}{(u+a)(1+a-u)},
\end{equation}
and $\dot u=u(1-u)[w(1-u)+lu+g(u)]$.
The integral $\int Q\,\dd t=\int_0^1Q\,\dd u/\dot u$ is finite and, for $w=l$, equals $\mathcal I(a)$ in Eq.~\eqref{eq:Ia}. By symmetry the three edges contribute equally, and $\Delta\ln\epsilon=3\mathcal I(a)$ per revolution, so that its sign determines attraction or repulsion. The construction requires $w(1-u)+lu+g(u)>0$, i.e., no equilibria on the edges, a condition that holds for $w=l>G(a)$.

\section{Lyapunov coefficients}
\label{app:l1}

We write Eq.~\eqref{eq:rep} for RPS as a planar system $\dot{\mathbf y}=\mathbf F(\mathbf y)$ in the coordinates $\mathbf y=(x_1,x_2)$, with $x_3=1-x_1-x_2$, and set $w-l=\delta_H(a)$. At the barycenter the Jacobian $J$ has eigenvalues $\pm i\omega$, with $\omega=\sqrt3(w+l)/6$. Let $\mathbf q$ and $\mathbf p$ satisfy $J\mathbf q=i\omega\mathbf q$, $J^{\!\top}\mathbf p=-i\omega\mathbf p$, $|\mathbf q|=1$, and $\langle\mathbf p,\mathbf q\rangle=\bar{\mathbf p}^{\!\top}\mathbf q=1$, and let $B$ and $C$ be the multilinear forms of the second and third derivatives of $\mathbf F$ at the barycenter. The first Lyapunov coefficient is~\cite{Kuznetsov2004}
\begin{equation}
\begin{split}
\ell_1=\frac{1}{2\omega}\,\mathrm{Re}\,\big\langle\mathbf p,\;&C(\mathbf q,\mathbf q,\bar{\mathbf q})-2B\big(\mathbf q,J^{-1}B(\mathbf q,\bar{\mathbf q})\big)\\
&+B\big(\bar{\mathbf q},(2i\omega-J)^{-1}B(\mathbf q,\mathbf q)\big)\big\rangle .
\end{split}
\end{equation}
All the quantities are rational functions of $a$, $w+l$, and $\sqrt3$, and a computer-algebra evaluation gives Eq.~\eqref{eq:l1}. With another normalization of $\mathbf q$ the prefactor changes, but the factor $2-3a$, and with it the Bautin point, does not.

At $B$ we obtained the sign of the second Lyapunov coefficient from a polynomial Lyapunov function of degree six, computed in exact arithmetic, which gives $\ell_2<0$.




\end{document}